\documentclass[a4paper]{article}

\usepackage[margin=2.5cm]{geometry}
\usepackage{amsmath}
\usepackage[]{amssymb}
\usepackage[]{mathrsfs}
\usepackage[]{eucal}
\usepackage[]{tensor}
\usepackage[]{amsthm}
\usepackage[]{bm}

\newcommand{\pf}[1]{\mathbf{#1}}
\newcommand{\dd}{\partial}
\newcommand{\hdg}{\star} 
\newcommand{\df}{\mathrm{d}}
\newcommand{\w}{\wedge}

\newcommand{\qqd}{\ , \quad}
\newcommand{\bc}{\begin{center}}
\newcommand{\ec}{\end{center}}
\newcommand{\be}{\begin{equation}}
\newcommand{\ee}{\end{equation}}

\newcommand{\F}{\pf{F}}

\newcommand{\FF}{\mathcal{F}}
\newcommand{\GG}{\mathcal{G}}
\newcommand{\HH}{\mathcal{H}}
\newcommand{\LL}{\mathscr{L}}
\newcommand{\JJ}{\mathcal{J}}
\newcommand{\KK}{\mathcal{K}}

\newcommand{\defeq}{\mathrel{\mathop:}=}

\usepackage{dsfont}
\newcommand{\nn}{\mathds{N}}

\newcommand{\rr}{\mathds{R}}

\usepackage[unicode]{hyperref}
\usepackage{xcolor}
\definecolor{pastgreen}{HTML}{669900}
\definecolor{pastblue}{HTML}{336699}
\definecolor{pastred}{HTML}{990000}
\definecolor{linkcol}{HTML}{663333}
\hypersetup{colorlinks,linkcolor={pastblue},citecolor={pastblue},urlcolor={pastblue}}

\begin{document}

\begin{flushright}
\texttt{ZTF-EP-23-08}
\end{flushright}

\vspace{20pt}

\bc
{\LARGE \textbf{Lagrangian Reverse Engineering for}}

\medskip

{\LARGE \textbf{Regular Black Holes}}

\vspace{20pt}

{\large Ana Bokuli\'c$^a$, Edgardo Franzin$^b$, Tajron Juri\'c$^c$ and Ivica Smoli\'c$^a$}

\bigskip

$^a$ Department of Physics, Faculty of Science, University of Zagreb, Bijeni\v cka cesta 32, 10000 Zagreb, Croatia

$^b$ Dipartimento di Fisica, ``Sapienza'' Universit\`a di Roma, piazzale Aldo Moro 5, 00185, Rome, Italy

$^c$ Rudjer Bo\v skovi\'c Institute, Bijeni\v cka cesta 54, HR-10002 Zagreb, Croatia
\ec

\vspace{30pt}

\textbf{Abstract.} Nonlinear extensions of  classical Maxwell's electromagnetism are among the prominent candidates for theories admitting regular black hole solutions. A quest for such examples has been fruitful, but mostly unsystematic and littered by the introduction of physically unrealistic Lagrangians. We provide a procedure which admits the reconstruction of a nonlinear electromagnetic Lagrangian, consistent with the Euler--Heisenberg Lagrangian in the weak-field limit, from a given metric representing a regular, magnetically charged black hole. 

\bigskip


\bigskip

\section{Introduction} 

The ubiquitous presence of singular spacetimes in general relativity \cite{TM1,TM2,TM3,ES,Ger} is usually taken as a signal of a certain incompleteness of the theory. Although it is expected that singularities will be absent in the underlying quantum theory of gravitation or the corresponding effective field theory counterpart \cite{KP22}, there is an intriguing possibility that spacetime may be ``regularized'' at energies well below those at which quantum gravitational effects become prominent. This motive is one of the many threads in the extensive quest for regular black holes \cite{CCZ18,SZ22,Maeda22,LYGM23,BES22}. If we set aside invocation of novel fields and interactions, arguably the most conservative approach is to turn to modifications of the classical Maxwell's electromagnetism, coming either from low-energy effective quantum field theory \cite{Dunne04}, e.g.\ the Euler--Heisenberg electromagnetic Lagrangian \cite{HE36}, or theoretical constructions motivated by symmetry arguments, e.g.\ the Born--Infeld model \cite{Born34,BI34} and the recently proposed ModMax theory \cite{BLST20}. Unfortunately, neither of these three theories leads to regular black hole solutions \cite{GSP84,SGP87,YT00,RWX13,FAGMLM21}. On the other hand, over the past several decades, we have witnessed a deluge of proposed electromagnetic theories \cite{Plebanski70,Sorokin21,Bronnikov22rev}, some of which apparently manage to de-singularize black holes. Nevertheless, all such proposals should be taken with a grain of salt and put under scrutiny, as most of them fail to meet some of the basic physical requirements which will be discussed below.

\smallskip

Let us consider nonlinear electromagnetic (NLE) theories based on a Lagrangian density $\LL(\FF,\GG)$ which is a function of two electromagnetic invariants, $\FF \defeq F_{ab} \, F^{ab}$ and $\GG \defeq F_{ab} \, {\hdg F}^{ab}$. This is the most general Lagrangian density constructed from the electromagnetic tensor $F_{ab}$, the spacetime metric $g_{ab}$ and the Levi--Civita tensor $\epsilon_{abcd}$, but it does not include theories with higher derivatives of $F_{ab}$ or nonminimal coupling to gravitation. We sort NLE theories into a simpler, so-called $\FF$-class for which $\LL = \LL(\FF)$, and a more general $\FF\GG$-class for which $\LL = \LL(\FF,\GG)$. Furthermore, it is convenient also to introduce the quantity $\HH \defeq (\FF^2 + \GG^2)^{1/2}$. Special attention is dedicated to the properties of theories for weak fields, where we have abundance of experimental insights. We say that a NLE theory respects the Maxwell's weak field (MWF) limit if $\LL = -\FF/4 + o(\HH)$ or the QED weak field (QEDWF) limit if $\LL = -\FF/4 + \kappa(4\FF^2 + 7\GG^2) + o(\HH^2)$, with $\kappa = \alpha^2/(360\,m_e^4)$, $\alpha$ being the fine-structure constant and $m_e$ the electron mass, in some neighbourhood of the origin of the $\FF$-$\GG$ plane. 

\smallskip 

One of the central objectives in this subfield of research is to find a black hole spacetime, regular in some sense (e.g.~having well-behaved curvature scalars and possibly being geodesically complete \cite{Carballo-Rubio:2019fnb}), which is a solution of the Einstein-NLE field equations, with an NLE Lagrangian satisfying the QEDWF limit. In addition, one might ask from the proposed NLE theory a possibility of fitting (with a proper choice of free parameters) to even higher orders of quantum corrections in the weak field limit. Unfortunately, as far as we are aware of, this goal has not yet been reached, although there are some candidates which come pretty close to it. 

\smallskip

The seminal paper which stirred the activity around NLE theories, the Ay{\'o}n-Beato--Garc{\'i}a's solution \cite{ABG98}, has a serious drawback as it tacitly uses different Lagrangians for different parts of the spacetime \cite{Bronnikov00}. On the other hand, their NLE interpretation \cite{ABG00} of the Bardeen's metric \cite{Bardeen68} does not even satisfy the MWF limit. Similar objections may be addressed to numerous other examples of regular black holes with NLE fields that have appeared in the literature (e.g.~some Lagrangians have a pole \cite{Kruglov14} or are not well-behaved \cite{Kruglov22} at the origin of the $\FF$-$\GG$ plane, while others, even if they satisfy the MWF limit \cite{FW16}, fail at the QEDWF limit, etc.). An interesting $\FF\GG$-class NLE theory, introduced by \cite{DGS19} and re-examined in \cite{Bronnikov22rev} (note that $4F_{ab} F^{bc} F_{cd} F^{da} = 2\FF^2 + \GG^2$), admits a regular magnetic black hole solution, but it seems that the proper QEDWF limit is inconsistent with the condition for the existence of horizons.

\smallskip

An important stumbling block on this road was revealed by the Bronnikov's theorem \cite{Bronnikov00}, originally proven only for the $\FF$-class theories but recently generalized \cite{BJS22b} to $\FF\GG$-class theories, which asserts that an \emph{electrically charged} static, spherically symmetric black hole cannot have bounded curvature scalars at its center, given that the NLE Lagrangian satisfies the MWF limit. On top of this, there is a series of additional theorems \cite{BJS22b} (see also \cite{TST23}) which constraint such an endeavour even in the presence of a magnetic charge.

\smallskip

A possible way forward is to start from a metric with desired properties and then find  the corresponding NLE theory, a procedure which we refer to as ``Lagrangian reverse engineering''. An effective approach in the case of $\FF$-class theories was paved by Bronnikov \cite{Bronnikov00,Bronnikov17} and re-discovered by Fan and Wang \cite{FW16} (cf.~objections in \cite{Bronnikov17c}). Still, it is not quite evident how to systematically achieve the goals that we have set above. The aim of this paper is to partially fill this gap by providing a procedure which works at least for magnetically charged black holes.

\smallskip

\emph{Conventions and notation}. Apart from the notation introduced above, we use the abbreviations $\LL_\FF \defeq \dd_\FF \LL$, $\LL_\GG \defeq \dd_\GG \LL$, etc.\ for partial derivatives of the Lagrangian density. Throughout the paper we use the standard Bachmann--Landau's ``big $O$'' and ``small $o$'' notation. Also, for any $\sigma > 0$, we write $F(r) = O_\infty(r^{-\sigma})$ as $r \to \infty$ if $F^{(k)}(r) = O(r^{-(\sigma+k)})$ for all $k \in \nn_0$ as $r \to \infty$.

\section{Static spherically symmetric spacetime with NLE fields} 

For simplicity, we shall focus on static, spherically symmetric black holes (cf.~remarks in \cite{PT69,DARG10} and \cite{BS23}), with the metric of the form
\be
\df s^2 = -f(r) \, \df t^2 + \frac{1}{f(r)} \, \df r^2 + r^2 (\df\theta^2 + \sin^2\theta \, \df\varphi^2) \, ,
\ee
where
\be
f(r) = 1 - \frac{2m(r)}{r} \, .
\ee
The regularity of the spacetime shall be analysed through three basic curvature invariants, i.e.\ the Ricci scalar $R \defeq g^{ab} R_{ab}$, the ``Ricci squared'' $S \defeq R_{ab} R^{ab}$, and the Kretsch\-mann scalar $K \defeq R_{abcd} R^{abcd}$, which for our metric read
\begin{align*}
R & = \frac{2}{r^3} \, (r^2 m')' \, \\
S & = \frac{2}{r^4} \, (4m'^2 + r^2 m''^2) \, \\
K & = \frac{4}{r^6} \, \big( 4(3m^2 - 4rmm' + 2r^2m'^2) + (4m - 4rm' + r^2 m'') r^2 m''\big) \, .
\end{align*}
We assume that the electromagnetic field inherits spacetime symmetries \cite{BGS17} and the corresponding ansatz reads
\be
\F = - E_r(r) \, \df t \w \df r + B_r(r) r^2 \sin^2\theta \, \df \theta \w \df \varphi \, .
\ee
From here, it is straightforward to evaluate the electromagnetic invariants,
\be
\FF = 2(B_r^2 - E_r^2) \qqd \GG = 4E_r B_r \, .
\ee
The generalized Maxwell's equations (gMax) may be put in the form \cite{BJS22b}
\be\label{eq:gMax}
B_r = \frac{P}{r^2} \qqd \LL_\FF E_r - \LL_\GG B_r = -\frac{Q}{4r^2} \, ,
\ee
where $P$ and $Q$ are the magnetic and electric charges of the black hole.
The assumed symmetries lead to the reduction of the Einstein's field equations to the system of differential equations
\begin{align}
-\frac{m'(r)}{r^2} & = \LL - \frac{QE_r}{r^2} \, \\
-\frac{m''(r)}{2r} & = \LL - \frac{4P^2}{r^4} \, \LL_\FF - \LL_\GG \GG \, .
\end{align}
Have we proposed a theory with a given $\LL$, the next step would be to find the ``mass function'' $m(r)$ and the electric field $E_r(r)$. However, we shall attack the problem from the opposite direction.

\section{Reverse engineering} 

Assuming that one has chosen an appropriate mass function $m(r)$, with the corresponding regular curvature invariants, we may ask whether it is possible to find an NLE Lagrangian density $\LL$ (together with an electric field $E_r$), such that the field equations are satisfied and $\LL$ obeys the weak field limits defined above. In general, this is a quite formidable task and, despite being mathematically exact, reverse engineering will be usually fraught with ambiguities. Namely, one has aim to reconstruct a Lagrangian $\LL(\FF,\GG)$ on some domain of the $\FF$-$\GG$ plane (if possible, on the whole plane).
However, the invariants of the electromagnetic field which is a solution of the Einstein--gMax equations typically cover only some smaller subset of the $\FF$-$\GG$ plane. Therefore, the strategy is either to (a) reconstruct $\LL$ as much as possible from one solution and then extrapolate the rest of the function (relying on experimental hints or some theoretical principles), or (b) reconstruct $\LL$ from two or more different classes of solutions. We shall follow the first option, with a couple of small tricks.

\smallskip

The first simplification is that we look at a special class of NLE Lagrangians of the form
\be
\LL(\FF,\GG) = \JJ(\FF) + \KK(\GG) \, .
\ee
In order to satisfy the QEDWF limit, we further assume that
\begin{align*}
\JJ(\FF) & = -\frac{1}{4} \, \FF + 4\kappa \FF^2 + o(\FF^2) \ \ \textrm{as} \ \ \FF \to 0 \quad \textrm{and} \\
\KK(\GG) & = 7\kappa \GG^2 + o(\GG^2) \ \ \textrm{as} \ \ \GG \to 0 \, .
\end{align*}
In particular, we assume that $\KK_\GG(0) = 0$. The second simplification is that we focus on the special, purely magnetic case, with  $Q = 0$ and $E_r = 0$. As an immediate consequence we have that $\GG = 0$ and $\LL_\GG = 0$, so that the gMax equation \eqref{eq:gMax} is automatically satisfied and the Einstein's field equations are reduced to
\begin{align}
-\frac{m'(r)}{r^2} & = \JJ(\FF) \, , \\
-\frac{m''(r)}{2r} & = \JJ(\FF) - \frac{4P^2}{r^4} \, \JJ_\FF(\FF) \, .
\end{align}
Now, as $\FF = 2B_r^2$, that is $r = (2P^2/\FF)^{1/4}$, it follows that the on-shell substitution
\be\label{eq:ersatz}
\JJ(\FF) = -\frac{m'(r)}{r^2} \Big|_{r = (2P^2/\FF)^{1/4}}
\ee
may recover the sought function $\JJ$. A nice feature of this procedure is that the other equation is automatically satisfied. In this way the problem is essentially reduced to the well-known reverse engineering for $\FF$-class theories \cite{Bronnikov00,FW16}, but we shall upgrade it to meet more stringent demands that we have set from the beginning.

\smallskip

To summarize, we search for a function $m(r)$ such that
\begin{itemize}
\item[(a)] the invariants $R$, $S$ and $K$ are bounded. 

This may be achieved if, for example, there is a real constant $M$ such that $m(r) = M + O_\infty(r^{-1})$ as $r \to \infty$ and $m(r) = O(r^3)$, $m'(r) = O(r^2)$ and $m''(r) = O(r)$ as $r \to 0$.

\item[(b)] The substitution \eqref{eq:ersatz} generates a Lagrangian density with the proper QEDWF limit.

As $\lim_{r\to\infty} \FF = 0$, this condition may be achieved if
\begin{equation*}
\frac{m'(r)}{r^2} = \frac{\beta_1}{r^4} + \frac{\beta_2}{r^8} + o(r^{-8}) \quad \textrm{as} \quad r \to \infty
\end{equation*}
with some appropriate real constants $\beta_1$ and $\beta_2$.

\item[(c)] The solution represents a black hole spacetime.

The existence of a black hole horizon follows if $f(r)$ has at least one zero on $\left< 0,+\infty \right>$.
\end{itemize}

\smallskip

In addition, given that the metric asymptotically approaches the magnetic Reiss\-ner--Nord\-str\"om metric,
\be
f(r) = 1 - \frac{2M}{r} + \frac{P^2}{r^2} + o(r^{-2}) \quad \textrm{as} \quad r \to \infty \, ,
\ee
we may identify $M$ and $P$ as the black hole mass and magnetic charge, respectively. Note that, strictly speaking, as $\FF > 0$ by construction in the magnetic case, this procedure allows us to reconstruct $\JJ(\FF)$ only for positive $\FF$, but we shall propose a ``natural'' extension of the function $\JJ$.

\section{Rational solution} 

It is expected that the problem defined above has infinite, albeit nontrivial solutions. We choose our candidate out of an arguably simpler family, a rational function
\be
2m(r) = \frac{2M - P^2 r^{-1} + sr^{-5}}{1 + br^{-8}} \, ,
\ee
with some positive parameters $s$ and $b$. Note that this ansatz has been built around the classical magnetic Reissner--Nordstr\"om solution, $2m_{\mathrm{RN}}(r) = 2M - P^2/r$, which was then expanded by adding carefully chosen powers of the radius in the numerator and denominator. First of all, we have $f(r) = 1 - 2M/r + P^2/r^2 + O(r^{-6})$ as $r \to \infty$ and the corresponding invariants are well behaved at the center\footnote{We refrain from writing out the complete expressions for $R$, $S$ and $K$; we only note that they are rational functions with a denominator of the form $(r^8 + b)^n$, where $n=3$ for $R$ and $n=6$ for $S$ and $K$.},
\be
R = \frac{12 s}{b} + O(r^4) \qqd S = \frac{36 s^2}{b^2} + O(r^4) \qqd K = \frac{24 s^2}{b^2} + O(r^4)
\ee
as $r \to 0$. In addition, all three invariants are bounded for $r \in \left[ 0,\infty \right>$ and fall-off as $O(r^{-6})$ or faster at infinity. The Reissner--Nordstr\"om limit $(s,b) \to (0,0)$ brings us back to $R = 0$ and unbounded $S = O(r^{-8})$ and $K = O(r^{-8})$ as $r \to 0$, but this limit may not be applied directly to the Taylor series above (just as e.g.~the limit $b\to 0$ applied to the Taylor series of the function $\psi(r) = 1/(r+b)$ around $r=0$ does not lead to the associated Laurent series $1/r$).

\smallskip

Black hole horizons appear at roots of the polynomial equation $f(r) = 0$, that is
\be
h(r) = r^8 - 2Mr^7 + P^2 r^6 - s r^2 + b = 0 \, .
\ee
First, note that $h(0) = b > 0$. Furthermore, a convenient way to write the polynomial $h$ is
\be
h(r) = r^6(r - r_-)(r - r_+) + b (1 - (r/r_0)^2)
\ee
with $r_\pm = M \pm \sqrt{M^2 - P^2}$ and $r_0 = \sqrt{b/s}$. Assuming that $M > P$ and $r_0 < r_+$, the intersection $V \defeq ( \left< r_-,r_+ \right> \cap \left< r_0,+\infty \right> ) \subseteq \left< 0,+\infty \right>$ is nonempty and for any $r \in V$ we have $h(r) < 0$. Thus, by the intermediate value theorem, the polynomial $h$ has (at least) one zero on $\left< 0,+\infty \right>$, implying that (at least) one black hole horizon exists. In addition, as $\lim_{r\to\infty} h(r) = +\infty$, we know that at least one more zero of $h$ (thus, one more horizon) exists.

\smallskip

The thermodynamic properties of rotating black holes with NLE fields in $\FF\GG$-class theory were analysed in \cite{BJS21} in complete generality, via covariant phase space formalism (the generalized Smarr formula was proven earlier in \cite{GS18}) and we shall not repeat the straightforward application of these formulas to this particular case.

\smallskip

Let us now turn to the central problem, i.e.\ the reconstruction of the Lagrangian. Taking the Taylor series with remainder of Eq.~\eqref{eq:ersatz}, we have
\be
\JJ(\FF) = -\frac{1}{4} \, \FF + \frac{5s}{8P^4} \, \FF^2 + O(\FF^{11/4})
\ee
as $\FF \to 0^+$. Thus, the QEDWF limit will be satisfied by taking $s = 32\kappa P^4/5$. It is convenient to write the other parameter as $b = \kappa \ell^2 P^4$, where $\ell$ is some dimensional constant with dimension of length. This leads us to the final form of the function $\JJ$,
\be\label{eq:JF}
\JJ(\FF) = -\frac{1}{5\left(4+\kappa(\ell\FF)^2\right)^2} \, \big( 20\FF - 320\kappa\FF^2 + \gamma\FF^{11/4} - 35\kappa\ell^2\FF^3 + 48\kappa^2\ell^2\FF^4 \big) \, ,
\ee
with $\gamma = 80 \sqrt[4]{2} \cdot \kappa\ell^2 MP^{-3/2}$. Now that the $\FF$-dependent part of the NLE Lagrangian has been reconstructed, one should ``read everything backwards'' and look at the Lagrangian parameters ($\kappa$, $\gamma$ and $\ell$ in Eq.~\eqref{eq:JF}) as our starting point, part of the definition of the theory. Here we notice that the mass $M$ and the magnetic charge $P$ are \emph{not independent} in this theory, a feature known in many other NLE theories\footnote{We note in passing that mass-charge dependence in absent in somewhat different approach to regular black holes with electromagnetic field nonminimally coupled to gravity \cite{CM20,CM21}.} \cite{Bronnikov22rev}. This fact can be interpreted at least in two ways, either (a) the mass $M$ is completely of electromagnetic origin, given by $M = \gamma P^{3/2}/(80 \sqrt[4]{2} \cdot \kappa\ell^2)$, or (b) the allowed magnetic charge is given by $P^{3/2} = 80 \sqrt[4]{2} \cdot \kappa\ell^2 M/\gamma$. An obstacle for the extension of the Lagrangian to negative values of $\FF$ appears with the noninteger power in the $\FF^{11/4}$ term, but may be resolved if we replace $\FF^{11/4}$ with $\mathrm{sgn}(\FF)|\FF|^{11/4}$. In this way the function $\JJ$ will be at least of class $C^2$ on the domain $\FF \in \rr$.

\smallskip

It is known that for NLE \cite{Plebanski70,BJS21} the null energy condition (NEC) holds iff $\LL_\FF \le 0$, while the dominant energy condition (DEC) holds iff $\LL_\FF \le 0$ and $g^{ab} T_{ab} \le 0$. As in our theory, for this particular metric, we have $\LL_\FF = -7r^8/(4b) + O(r^9)$ and $g^{ab} T_{ab} = -48/(5\pi\ell^2) + O(r^4)$ as $r \to 0$, we know that NEC and DEC hold at least in some neighborhood of the center $r = 0$ (cf.~remarks in \cite{BV14a,BV14b}).

\smallskip

Now, the important question is whether this metric ansatz may be extended in order to admit fitting of the associated function $\JJ$ to even higher orders of some established corrections to Maxwell's electromagnetic Lagrangian. We note that the generalized function
\be\label{genAnsatz}
2m(r) = \frac{2M - P^2 r^{-1} + s_1r^{-5} + \dots + s_k r^{-(4k+1)}}{1 + br^{-4(k+1)}}
\ee
with real parameters $(s_1,\dots,s_k,b)$ asymptotically behaves as the magnetic Reissner--Norstr\"om black hole and has regular invariants at the center,
\be
R = \frac{12 s_k}{b} + O(r^4) \qqd S = \frac{36 s_k^2}{b^2} + O(r^4) \qqd K = \frac{24 s_k^2}{b^2} + O(r^4)
\ee
as $r \to 0$, given that $b \ne 0$. A technical obstacle in the analysis of the horizon structure for this metric is that we have to deal with a polynomial equation of high degree. Furthermore, the ansatz \eqref{genAnsatz} leads to a Lagrangian function $\JJ$, with the convenient form
\be
\JJ(\FF) = -\frac{1}{4} \, \FF + \frac{5s_1}{8P^4} \, \FF^2 + \dots + \frac{(4k+1)s_k}{2^{k+2} P^{2k+2}} \, \FF^{k+1} + O(\FF^{(4k+7)/4})
\ee
as $\FF \to 0$. The formal extension of the function $\JJ$ to negative values of $\FF$ may be performed as above.

\smallskip

As was stressed from the beginning, our construction rests upon the simplification that the black hole is purely magnetically charged, which ``hides'' any $\GG$-dependent terms in the Lagrangian. A next step would be the generalization of the NLE Lagrangian reconstruction from dyonic regular black holes. Noting that $\FF^2 + \GG^2 = 4(E_r^2 + B_r^2)^2$ and $\FF + \sqrt{\FF^2 + \GG^2} = 4P^2/r^4$, a possible Lagrangian candidate could be read off from one of the Einstein's field equations,
\be
\LL(\FF,\GG) = \frac{QE_r - m'(r)}{r^2} \Big|_{*}
\ee
where $*$ stands for the substitution $r = (4P^2/(\FF + \sqrt{\FF^2 + \GG^2}))^{1/4}$ and $E_r = \GG r^2/4P$. However, this equality holds only on-shell, for the particular solution, and there is no \emph{a priori} guarantee that this procedure reconstructs the \emph{general} corresponding Lagrangian. Also, a highly nontrivial consistency check is that the rest of the field equations are satisfied with such an ansatz (which is not automatic as in the purely magnetic case). We leave the investigation of this question for future work.

\section{Final remarks} 

We have outlined a programme focused on the quest for more realistic models of regular classical black holes, which do not invoke novel fields or extensions of general relativity, but are based upon nonlinear electromagnetic fields which agree with the weak field limits of quantum corrections to Maxwell's electromagnetic Lagrangian. 
Due to no-go theorems \cite{Bronnikov00,BJS22b}, the ``price'' to be paid here is the introduction of a magnetic charge and the crucial question is whether a small amount (consistent with astrophysical constraints) might be sufficient to ``de-singularize'' black holes.
Resting upon the elements of previously used reverse engineering procedures, our model establishes a first step in this quest. The next, important one would be the generalization to static regular dyonic black holes and then, even more ambitious, the generalization to rotating regular black holes (see \cite{TSA17} for an attempt to construct a regular rotating black hole with $\FF$-class NLE Lagrangian, alas without a functional form of the Lagrangian, and criticism of Newman--Janis algorithm in \cite{KTS22}). Finally, given that we have an example of rotating regular black hole, with an NLE Lagrangian passing all the criteria set above, one has to analyse its stability \cite{NYS20} and questions about ``mass inflation'' at the inner horizon \cite{CRDFLPV22,Franzin:2022wai} in order to corroborate the inclusion of such solution among physically realistic models.

\section*{Acknowledgements} 

Research of A.B., T.J.\ and I.S.\ was supported by the Croatian Science Foundation Project No.~IP-2020-02-9614.
E.F.\ acknowledges funding support from the MIUR PRIN (Grant 2020KR4KN2 ``String Theory as a bridge between Gauge Theories and Quantum Gravity''), FARE (GW-NEXT, CUP:\ B84I20000100001, 2020KR4KN2) programmes, and from EU Horizon 2020 Research and Innovation Programme under the Marie Sk\l{}odowska-Curie Grant Agreement no.~101007855.

\bibliographystyle{amsalpha}
\bibliography{reg}

\end{document}